\documentclass{emulateapj}

\usepackage{natbib}
\slugcomment{ApJ accepted}

\shorttitle{Identification of a $z=4$ Submillimeter Galaxy}
\shortauthors{Knudsen et al.}

\begin{document}


\title{Physical properties and morphology of a newly identified compact
z=4.04 lensed submillimeter galaxy in Abell 2218}


\author{Kirsten~K.~Knudsen\altaffilmark{1}, Jean-Paul~Kneib\altaffilmark{2,3}, Johan Richard\altaffilmark{4,3}, Glen Petitpas\altaffilmark{5}, Eiichi~Egami\altaffilmark{6}}
\altaffiltext{1}{Argelander-Institut f\"ur Astronomie, Auf dem H\"ugel 71, D-53121 Bonn, Germany; knudsen@astro.uni-bonn.de}
\altaffiltext{2}{Laboratoire d'Astrophysique de Marseille, OAMP, Universit\'e, Aix-Marseilles \& CNRS, 38 rue Fr\'ed\'eric Joliot Curie, 13388 Marseille cedex 13, France}
\altaffiltext{3}{California Institute of Technology, MS 105-24, Pasadena, CA 91125, USA}
\altaffiltext{4}{Durham University, Physics and Astronomy Department, South
Road, Durham DH3 1LE, UK}
\altaffiltext{5}{Harvard-Smithsonian Center for Astrophysics, 60 Garden Street, Cambridge, MA 02138, USA}
\altaffiltext{6}{Steward Observatory, University of Arizona, 933 North Cherry Avenue, Tuscon, AZ 85721, USA} 


%\altaffiltext{1}{Visiting Astronomer, Cerro Tololo Inter-American Observatory.
%CTIO is operated by AURA, Inc.\ under contract to the National Science
%\altaffiltext{2}{Society of Fellows, Harvard University.}
%\altaffiltext{3}{present address: Center for Astrophysics,
    %60 Garden Street, Cambridge, MA 02138}


\begin{abstract}
We present the identification of a bright submillimeter (submm) source,
SMM\,J163555.5+661300, detected in 
the lensing cluster Abell~2218, 
for which we have accurately determined the position using observations from
the Submillimeter Array (SMA).  
The identified optical counterpart has a spectroscopic redshift of $z =
4.044\pm0.001$ if we attribute the single emission line detected at
$\lambda=6140$\AA\ to Lyman-$\alpha$.  This redshift identification is in
good agreement with the optical/near-infrared photometric redshift as well as
the submm flux ratio $S_{450}/S_{850} \sim 1.6$, the radio-submm flux
ratio $S_{1.4} / S_{850} < 0.004$, 
and the $24\mu$m to 850$\mu$m flux ratio $S_{24}/S_{850} < 0.005$.
Correcting for the gravitational lensing
amplification of $\sim 5.5$, we find that the source has a far-infrared
luminosity of $1.3\times10^{12}$\,L$_\odot$, which implies a star formation
rate of 230\,M$_\odot$\,yr$^{-1}$.  This makes it the 
lowest-luminosity SMG known at $z>4$ to date.  
Previous CO(4-3) emission line obserations yielded a non-detection, for which
we derived an upper limit of the CO line luminosity of $L_{\rm CO}^{'} =
0.3\times10^{10}$\,K\,km/s\,pc$^{-2}$, which is not inconsistent with the
$L^{'}_{\rm CO} - L_{FIR}$ relation for starburst galaxies.  
The best fit model to the optical and
near-infrared photometry give a stellar population with an age of 1.4\,Gyr
and a stellar mass of $1.6\times10^{10}$\,M$_\odot$.   The optical morphology
is compact and in the source plane the galaxy has an extent of $\sim6\times
3$\,kpc with individual star forming knots of $<500$\,pc in size.  
J163556 is not resolved in the SMA data and we place a strict upper limit on the
size of the starburst region of $8{\rm kpc} \times 3{\rm kpc}$, which
implies a lower limit on the star formation rate surface density of
12\,M$_\odot$\,yr$^{-1}$\,kpc$^{2}$.  
The redshift of J163556 extends the redshift distribution of faint, lensed
SMGs, and we find no evidence 
that these have a different redshift distribution than bright SMGs.  
\end{abstract}


%% subject header.  Objects should be in the appropriate "individual"
%% headers (e.g. quasars: individual, stars: individual, etc.) with the
%% additional provision that the total number of headers, including each
%% individual object, not exceed six.  The \objectname{} macro, and its
%% alias \object{}, is used to mark each object.  The macro takes the object
%% name as its primary argument.  This name will appear in the paper
%% and serve as the link's anchor in the electronic edition if the name
%% is recognized by the data centers.  The macro also takes an optional
%% argument in parentheses in cases where the data center identification
%% differs from what is to be printed in the paper.

\keywords{galaxies: individual (SMM\,J163556+661300) --- galaxies: high-redshift --- submillimeter}


\section{INTRODUCTION}

The redshift distribution of submillimeter galaxies (SMGs) is vital to our
understanding of the nature of these galaxies and their role in general
picture of galaxy formation and evolution.  The first systematic
identification and redshift determination of SMGs relied on the radio
counterparts and was carried out by \citet{chapman03,chapman05}.  While the
radio and optical/near-infrared counterparts have 
much better determined positions compared to the original $15''$ resolution
submillimeter observations and thus is an efficient approach for identifying
the underlying galaxy, the radio emission rapidly dims with the distance and
thus is likely to introduce a bias against SMGs with redshift $z>3.5$.   
Due to the negative {\em K}-correction (sub)mm
observations offer a great opportunity to search for galaxies beyond
$z>4$ and one would expect to find SMGs up to redshifts $z\sim6-7$.
Through modeling of the SMG population \citet{chapman05} predict that about
10\% of the population is at redshift $z\geq 4$.  
\citet{wall08} suggested a 
bimodal redshift distribution, where the brightest SMGs dominate at the
highest redshifts and the fainter SMGs dominate towards lower redshifts,
as initially proposed by \citet{ivison02}, 
equivalent to 'cosmic down-sizing'.  

To date, there has been no systematic samples of very high redshift ($z>3.5$)
SMGs, however, \citet{dannerbauer02,dannerbauer04} and \citet{younger07} have
presented several high redshift candidates that were identified through
(sub-)mm interferometry (with and without radio counterparts).  Three SMGs with
accurate redshifts are \objectname{COSMOS J100054+023436}
\citep[$z=4.547$;][]{capak08,schinnerer08} and \objectname{GN20} and \objectname{GN20.2}
\citep[both $z=4.05$;][]{daddi09}.  
These three sources have CO redshifts providing solid evidence for correct
identification.  
The galaxy \objectname{GOODS 850-5} (a.k.a.\ GN10) does not have an optical
spectroscopic redshift, but based on photometric redshifts two independent
groups conclude it is $z>4$, possibly $z\sim6$,
\citep{dannerbauer08,wang07,wang09}.  
In a recent paper, \citet{daddi09b} presented a redshift based on a single
millimeter wavelength emission line, which they interpret as the CO(4-3)
line placing the galaxy at redshift $z=4.0424$.  
Very recently, the highest redshift SMG known to date has been identified,
namely the $z=4.76$ galaxy LESSJ033229.4-275619 in the LABOCA map of E-CDF-S
\citep{coppin09}. 

In this paper we present the identification of the optical counterpart of the
submillimeter source \object{SMMJ163555.5+661300} (henceforth J163556), which
was detected in the deep SCUBA\footnote{Submillimetre Common-User Bolometer
Array \citep{holland99} mounted at the James Clerk Maxwell Telescope (JCMT)}
map of \object{Abell 2218} \citep{kneib04,knudsen06} with a lensed flux of
$S_{850} = 11.3$\,mJy.
It is important to note that this is a $10\sigma$ SCUBA detection, rendering
it a very solid detection and not a spurious source. 
Exploiting the strong gravitational lensing effect by the cluster potential,
these SCUBA observations have yielded the largest number of faint submm
galaxies detected in one field.  
Here we present multiwavelength followup and spectroscopic observations. 

Throughout we will assume an $\Omega=0.3$, $\Lambda=0.7$
cosmology with $H_0=70$\,km\,s$^{-1}$\,Mpc$^{-1}$. 
With this cosmology, at $z=4$ $1''$ corresponds to 6.95 kpc.

\section{OBSERVATIONS}

\subsection{Submillimeter interferometric observations} 

SMMJ163556 was observed at the Submillimeter Array \citep{ho04} on
Mauna Kea in compact configuration on UTC 05 May 2009 using 7 of the 8
SMA antennas. The resulting image created has $uv$-coverage from 13.7
k$\lambda$ to 121.6 k$\lambda$ at the observed local oscillator frequency of 
340 GHz. Observations were made in very dry conditions, with $\tau_{225}$
ranging from 0.04 to 0.06. The full 4 GHz (2 GHz in each sideband
separated by 10 GHz) were combined to achieve a total rms noise of 1.9
mJy in the final map.

The SMA data were calibrated using the MIR software package developed
at Caltech and modified for the SMA. Gain calibration was performed
using the nearby quasars 1642+689, 1638+573, and 1849+670. Absolute
flux calibration was performed using real-time measurements of the
system temperatures, with observations of Callisto, Ganymede, Uranus
and Neptune to set the scale. Bandpass calibration was done using
Callisto. The data were imaged using Miriad \citep{sault95}, where the
resulting synthesized beam is $2.1''\times1.5''$ with a position angle
P.A.$=39\deg$.  
To verify the astrometry, the position of 1638+573 was calibrated using
1642+689 as reference sources; 
both calibrators have accurately known positions known from the
International Celestial Reference Frame.  
A fit to the {\em uv}-data of 1638+573 shows that it deviates from the known
radio position by $\Delta$R.A.~$=0.14''$ ($0.10''$) and $\Delta$Dec.~$=0.02''$
($0.005''$) in the lower (and upper) sideband. 
As the distance between 1638+573 and 1642+689 is 11.6 degrees, four times
larger than that between 1642+689 and J163556, the fit provides a
conservative upper limit to the systematics uncertainties in the SMA
astrometry.

\subsection{Radio observations}

Radio maps of A2218 at 1.4\,GHz and 8.2\,GHz were presented 
by \citet{garrett05}.  The depth of the 1.4\,GHz data, which 
were obtained with the Westerbork Radio Synthesis Telescope, 
is 1$\sigma$ r.m.s.\ $\sim$15\,$\mu$Jy, and the depth of the 
Very Large Array 8.2\,GHz map is $\sim$6\,$\mu$Jy.  
J163556 is not detected in the two maps, placing 
3$\sigma$ upper limits of $S_{1.4} < 45$\,$\mu$Jy and 
$S_{8.2} < 18$\,$\mu$Jy. 

\subsection{Optical and near-infrared imaging}
\label{subsect:optnirphot}

High angular resolution space-based optical and near-infrared observations
have been obtained with the Advanced Camera for Surveys (ACS) and NICMOS on
the Hubble Space Telescope for the programs 9292, 9452, 9717, 10325, and
10504 with a total exposure time in the range from 1.6 to 5.5 hours.  The
data reduction of these data is discussed in detail in \citet{richard08}.  In
summary the ACS data were reduced using the {\tt multidrizzle} software
\citep{koekemoer02}, which also removes cosmic rays and bad pixels and
combines the dithered frames to correct for camera distortions.   The NICMOS
data were reduced following the procedures given in the NICMOS data reduction
handbook, then bad pixels were flagged, and subsequently using the {\tt
LACOSMIC} \citep{vandokkum01} IRAF procedures for cosmic ray rejection. 

Ground-based observations in the $K$-band were obtained using MOIRCS on
Subaru \citep[Multi-Object InfraRed Camera and Spectrograph;][]{ichikawa06}
as a part of the large search of extremely high redshift lensed galaxies
\citep{richard08}.  The integration time is $\sim$5 hr with good seeing
conditions (0.3$''$--0.4$''$) using dithered exposures of 50 seconds
duration.  The data were reduced using the MCSRED software
package\footnote{Available from
\url{http://www.naoj.org/staff/ichi/MCSRED/mcsred.html}.} and the details on this
is described in \citet{richard08}. 

The astrometry of the optical and near-infrared images was obtained using the
USNO and the 2MASS catalogs for the field and the dispersion gives an
accuracy $<0.2''$. 

\begin{figure}
\begin{center}
\vskip5cm
\includegraphics[width=9cm]{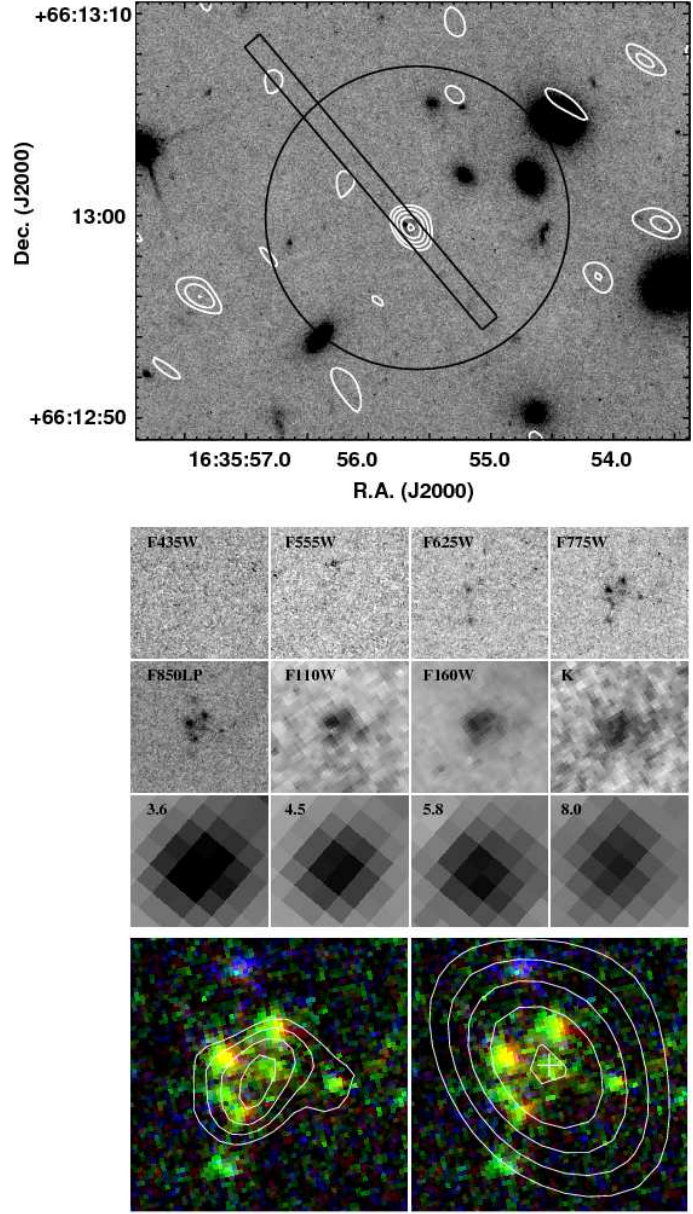} 
\caption{
{\em Top panel:}~ The F850LP ($z'$) image overlayed with the SMA contours
(white; in steps of 1$\sigma$ rms.\ starting at 2$\sigma$).  The black circle
shows the SCUBA beam (FWHM of 15$''$). 
The black rectangle shows the position of the slit used for optical
spectroscopy.  
{\em Middle panel:}~  
Mosaic of the optical and near-infrared images of J163556. The size of each
image is $\sim$4$''\times$4$''$.  
{\rm Bottom panel:}~ 
Is a $\sim2''\times2''$ true-color image of the F555, F775, and F850LP
overlayed with {\em left:} \ $K$-band contours and {\em right:} \ SMA
contours, where the plus sign indicates the submm position as measured with
the SMA (the plus is $0.2''\times0.2''$). 
\label{fig:phot}}
\end{center}
\end{figure}

Optical and near-infrared photometry 
was performed using SExtractor \citep{bertin96} for a 0.5 arcsec aperture for
the HST data and with a standard aperture correction in the $K$-band. 
When undetected, we place 3$\sigma$ upper limits. 
The photometry 
for the source z4, which is at a distance of 0.6 arcsec from the SCUBA
position, 
is given in Table \ref{tab:phot} and all magnitudes are in the
AB system \citep{oke74}. 
The images are shown in Fig.~\ref{fig:phot}.

\subsection{Mid-infrared imaging}

The ultra-deep (10 hr/band) IRAC data of Abell 2218 were obtained as
part of the Spitzer GO-2 program PID 20439 (PI: Eiichi Egami) while
the deep ($\sim$3000s) MIPS 24 $\mu$m data were taken as part of the
Spitzer GTO program PID 83 (PI: George Rieke).  See \citet{egami06}
for the details of the IRAC and MIPS data reductions.

J163556 is detected in all four IRAC bands (3.6, 4.5, 5.8 and 8.0
$\mu$m) but not at 24 $\mu$m.  A convolved HST image was used for
determining the HST-IRAC colors.  The convolved HST image was also used for
the astrometry registration. 
The MIPS non-detection sets a 3$\sigma$
upper limit of $<$ 40 $\mu$Jy at 24 $\mu$m.  J163556 is not detected
in the {\em Infrared Space Observatory} ISOCAM maps at 7 and 15 $\mu$m
\citep{metcalfe03}. We place 3$\sigma$ upper limits of 63 and 150
$\mu$m Jy, respectively.

\subsection{Optical spectroscopy}

We conducted deep multi-slit spectroscopy using the Low Resolution Imaging
Spectrograph \citep[LRIS; mounted at the 10m Keck Telescope on Mauna Kea,
Hawaii;][]{oke95} of sources in rich galaxy cluster field A\,2218.  The data
were obtained during the nights of June 30 and July 1, 2003, which had
reasonable seeing, $\sim 0.8\arcsec$, but were not fully photometric (with
some cirrus).  We obtained a crude flux calibration of our observations using
Feige 67 and 110 as spectrophotometric standard stars.
We observed the galaxy z4 (see Section \ref{subsect:optnirphot}) for a total of 11.8\,ks using 
the 400/8500 red grating and a total of 10.6\,ks using
the 600/4000 blue gratings offering respectively a
spectral dispersion of 0.63\AA\ pixel$^{-1}$ in the blue and 1.86\AA\
pixel$^{-1}$ in the red.  No flux is detected in the blue channel below
5600\AA, where the dichroic splitter is located. 
We reduced these data that was part of a multi-object mask using standard
IRAF scripts.

\section{RESULTS}

We fitted a point source to the SMA data to determine the position of the
J163556.  When fitting to the {\it uv}-data we find point source at the
position RA, Dec (J2000) = 16:35:55.67, 66:12:59.51 with positional errors of
0.15 arcsec with a flux of $10.7\pm2.0$\,mJy, and consistent results are
found for the upper and lower sidebands, individually.  When fitting to the
synthesized map, we find RA, Dec (J2000) = 16:35:55.64$\pm$0.02,
+66:12:59.44$\pm0.14$ and a flux of $10.7\pm0.2$\,mJy, with similar
results obtained using routines JMFIT from AIPS and imfit from MIRIAD.  
We note that this position is only an arcsecond from the SCUBA position
\citep{knudsen08}, demonstrating that good positions can be obtained for
sources in the coarse resolution maps when the signal-to-noise ratio is
sufficiently high (in this case $\sim 10$).  The flux determined by the SMA
data agrees with the SCUBA flux, further verifying the source is unresolved
in the SMA image.  

In Fig.~\ref{fig:phot} we show the SMA data overlaid on the optical data.
The multiwavelength photometry for the underlying counterpart galaxy of
J163556 is presented in Table~\ref{tab:phot}.   
The optical/near-infrared position is determined in filter bands F625W to
$K$.  Because of the complex morphology of z4, we smoothed the ACS images
with a $0.6''$ Gaussian kernel and the F110W band with a $0.5''$ Gaussian
kernel before determining the position.  The morphology varies between the
different optical/near-infrared bands.  We derive a mean
optical/near-infrared position for z4 of RA,Dec(J2000) =
16:35:55.67$\pm$0.03, +66:12:59.42$\pm$0.08.    
The offset between this mean position and the SMA position derived from the
{\em uv}-data is $0.09''$, and the uncertainty on the offset is $0.23''$ in
R.A.\ and $0.17''$ in Dec, hence we conclude that optical and submm position
agree well.  

\begin{figure*}
\begin{center}
\epsscale{.99}
\includegraphics[width=15cm]{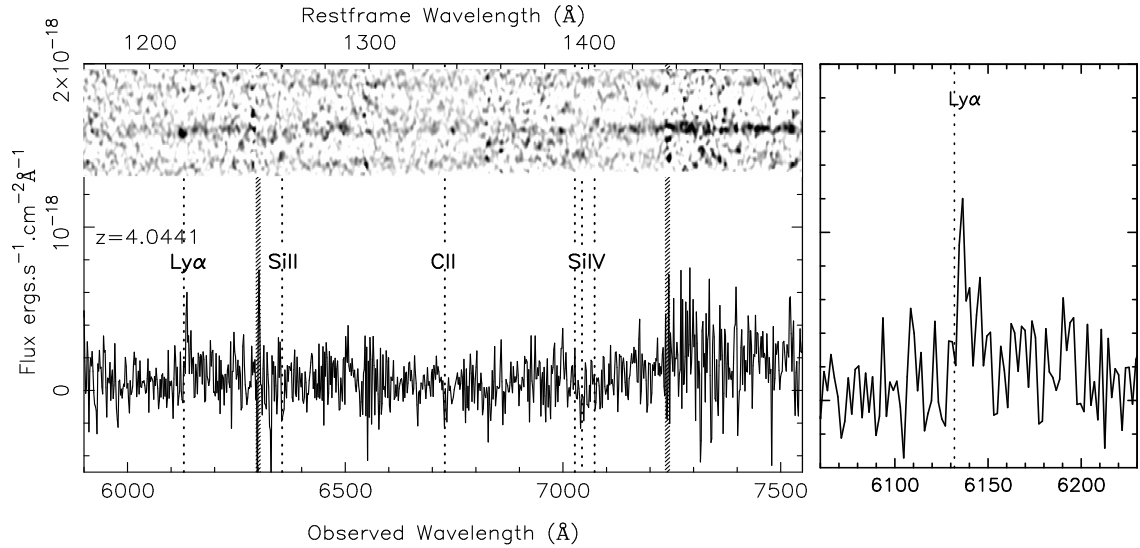}
\caption{Optical spectroscopic identification of J163556.  
(Top) The 2D sky subtracted spectrum that shows a faint emission line. 
It has been smoothed using a Gaussian kernel to visually enhance the emission.  
(Bottom) The 1D sky subtracted spectrum.  The hashed vertical lines 
indicate the location of bright sky lines.  The best fit to the Ly$\alpha$
emission as well as several (marginal) absorption lines yield a redshift of
$z=4.044\pm0.001$; the dotted vertical lines show the position of the
Ly$\alpha$ line and the absorption lines. The {\em right panel} shows a zoom
in on the Ly$\alpha$ line.  
\label{fig:spec}}
\end{center}
\end{figure*}

Figure  \ref{fig:spec} shows the 2D and 1D extraction
of this object within the 6000-6700\AA\ window, where we have identified an
(asymmetric) emission line centered at $\lambda=6135$\AA.  Two independent
reductions of the data were carried out producing similar results, and thus
the emission line is most probably not spurious.  A faint continuum
is weakly visible redwards of the detected emission line.   
In light of other evidence discussed below, we believe this line to be 
Lyman-$\alpha$, which correspond to a redshift
$z=4.044\pm 0.001$, obtained through fitting to the emission
line and a few absorption lines.  We note these absorption lines are marginal
and thus carry a low weight.  A break in the continuum bluewards of the line
as well as the lack of continuum emission in the blue part of the spectrum
supports this identification.  
Alternatives at low redshift such as [OIII] at $z=0.23$ or [OII] at $z=0.65$
seem less likely due to the absence of other lines and/or the lack of
continuum on the blue side.

\section{DISCUSSION} 

\subsection{Redshift}

\begin{figure}
\begin{center}
\hskip4mm\includegraphics[width=7.5cm,height=4.5cm]{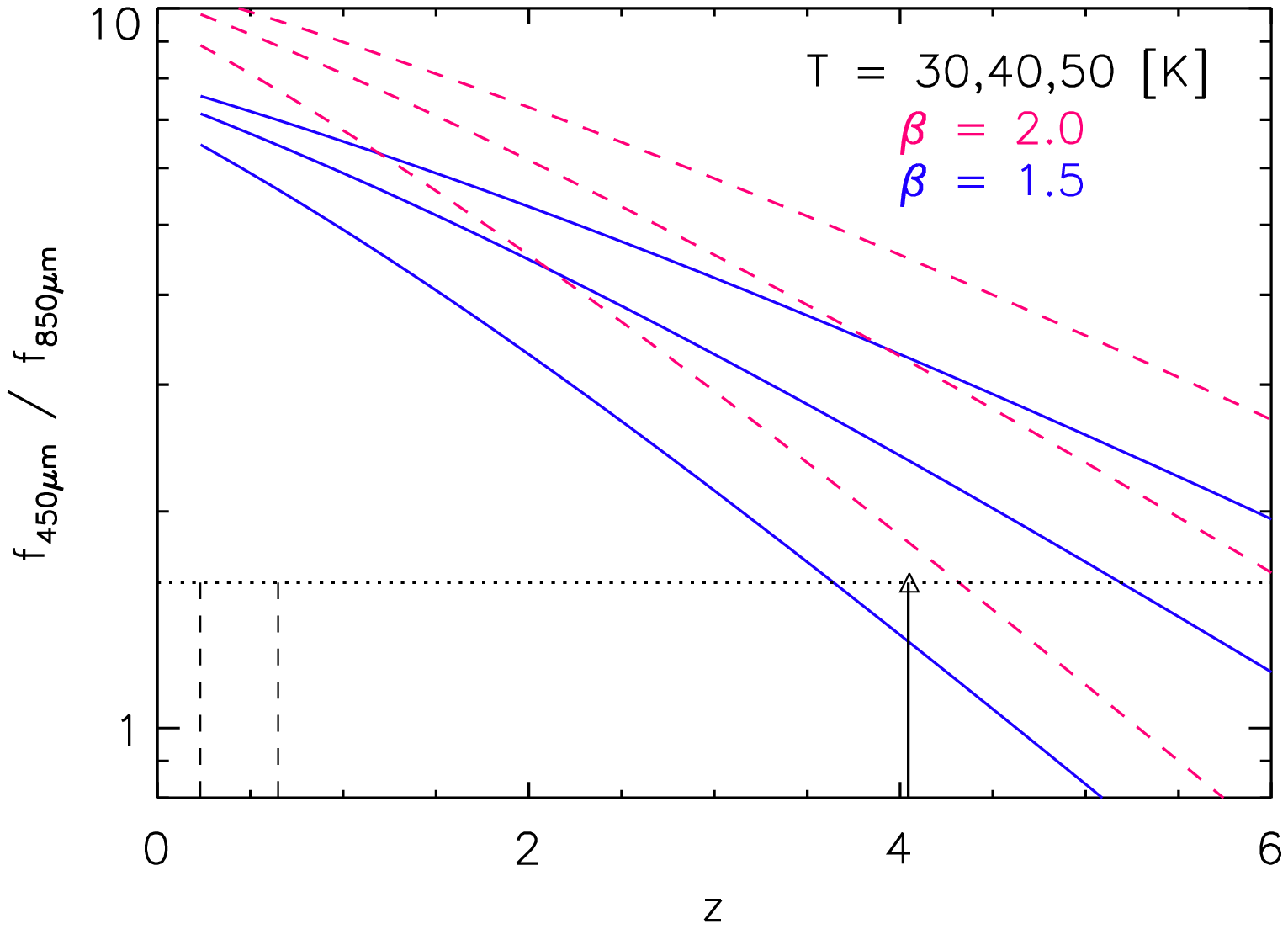} \\
\includegraphics[width=7.9cm,height=4.5cm]{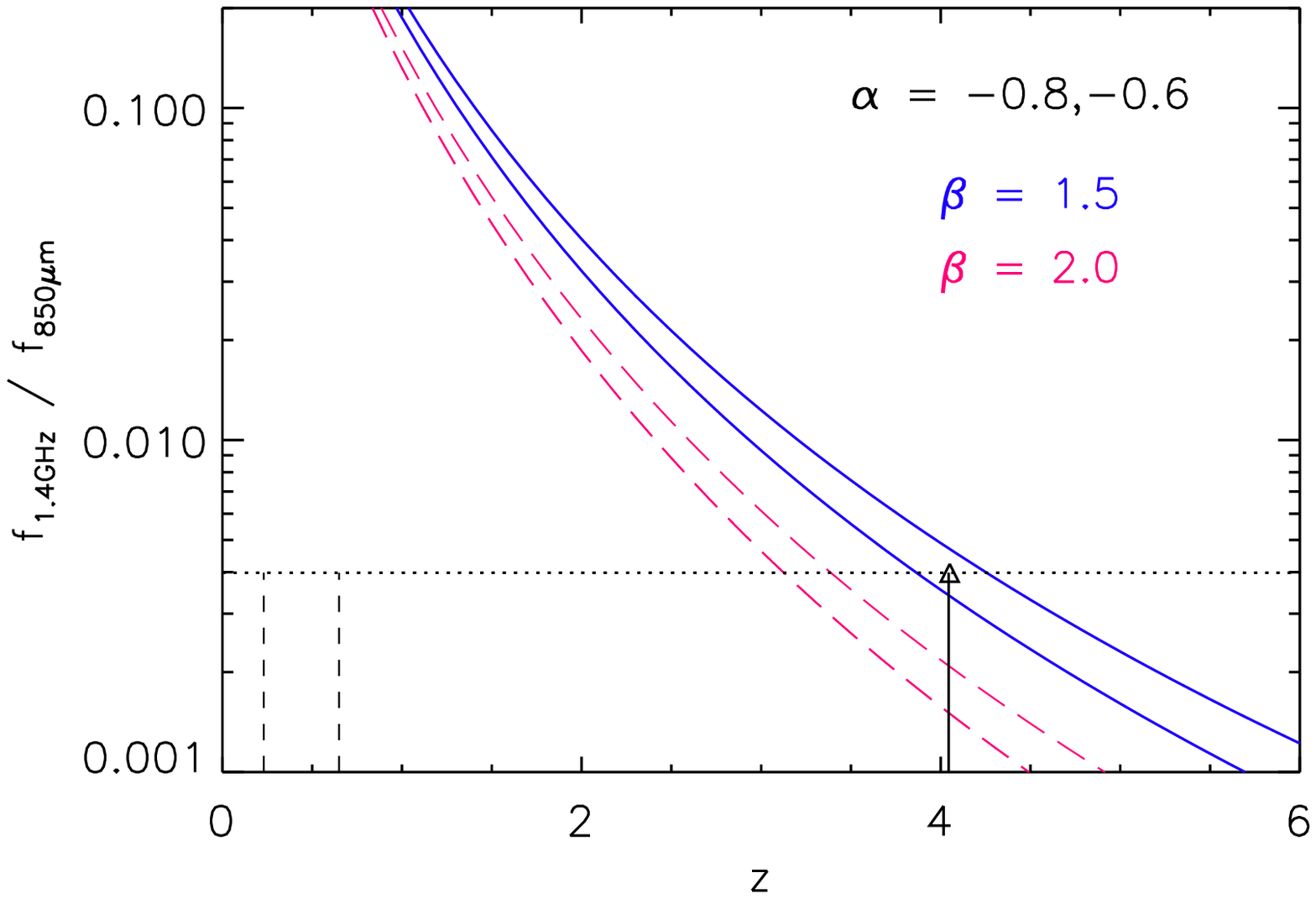}
\end{center}
\caption[]{The predicted 450$\mu$m-850$\mu$m ratio and 1.4GHz-850$\mu$m flux
ratio as function of redshift.  The former is calculated assuming a modified
blackbody SED (with $T$ going from 30 to 50K from left to right) and the
latter is based on the relation from \citet{carilliyun99} (with $\alpha =
-0.8$ as the left curve). 
The horizontal dotted line shows the flux ratio and flux ratio limit,
respectively.  The vertical solid line indicates $z=4.044$, while the
vertical dashed lines show the redshift 0.23 and 0.65. 
\label{fig:fluxratios}}
\end{figure}

The ratio of the 450\,$\mu$m and 850\,$\mu$m fluxes of 1.6 is strongly
indicative of J163556 being a high redshift source with $z \geq 3.5$,
assuming that the far-infrared spectral energy distribution (SED) is
well-described by a modified black body (as illustrated in
Fig.~\ref{fig:fluxratios}).  A high redshift is supported by the 1.4GHz to
850\,$\mu$m flux ratio of $<0.004$, where the depth of the radio data allows
a constraint of $z>3$.  In Fig.~\ref{fig:fluxratios} we show how the 1.4GHz
to 850$\mu$m flux density ratio changes with redshift using the relation from
\citet{carilliyun99}.  As discussed in \citet{blain99} and
\citet{carilliyun99}, the radio-submm flux can be used as a redshift
indicator, assuming that both the radio and submm emission is powered by star
formation.  
Additional evidence for J163556 being at high redshift ($z>3.5$) comes from
the $24\mu$m to $850\mu$m flux ratio $S_{24} / S_{850} < 0.005$
\citep[e.g.\ Fig.~3][]{pope06}. 
We note that the individual flux ratios in the radio, submm, and
far-infrared, only provide crude redshift estimates and are each known to
have a large scatter \citep[e.g.][]{carilli00}. 

\begin{figure}
\begin{center}
\includegraphics[width=8.5cm,height=7cm]{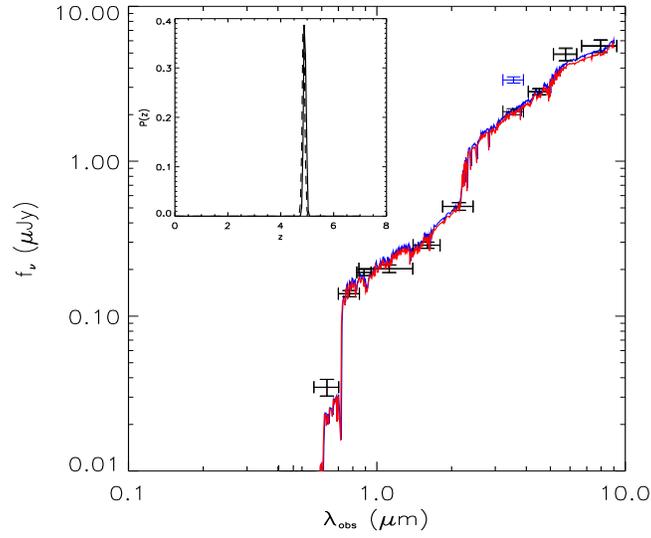}
\caption[]{The SED from the F435W to 8.0$\mu$m photometry overlayed with the
best-fit spectrum from {\it Hyperz} from the photo-$z$ modeling of the
optical counterpart of J163556.  The blue 3.6$\mu$m shows the photometry as
measured, and the black 3.6$\mu$m data point has been corrected for H$\alpha$
emission line contribution.  The blue spectrum shows the fit when including
the H$\alpha$ corrected 3.6$\mu$m data point ($z_{ph} = 4.77$) and the red spectrum shows the
fit when excluding the 3.6$\mu$m photometry from the photo-$z$ estimate
($z_{ph} = 4.67$). 
The insert shows the output probability distribution for the photometric
redshift from {\it Hyperz}, where the solid line corresponds to the fit
including the corrected 3.6$\mu$m photometry and the dashed line corresponds
to the fit with 3.6$\mu$m excluded. 
\label{fig:photz_ha}}
\end{center}
\end{figure}

As the optical spectrum has only a single emission line, we calculate the
photometric redshift using the {\it Hyperz} code \citep{hyperz} and the
photometry from the F435W ($=B$) band to the IRAC bands.    
We see an excess in the 3.6$\mu$m band (shown in Fig.~\ref{fig:photz_ha}),
which is most probably caused by H$\alpha$ emission \citep{yabe08}.  For
redshifts $z=3.84-5.03$ the H$\alpha$ line will shift through the 3.6$\mu$m
IRAC channel.
We make a correction for the H$\alpha$ emission deriving a line luminosity
based on 10\% of the star formation rate assuming a typical FIR luminosity
for SMGs. 
Excluding the 3.6$\mu$m data point, we find a photo-$z$
of 4.67, while including the H$\alpha$ corrected 3.6$\mu$m data point we find
a photo-$z$ of 4.77.  This high photometric redshift strongly supports our
identification of the emission line in the optical spectrum as Ly$\alpha$ and
not e.g.\ low redshift [OII]. 

\begin{figure}
\begin{center}
\includegraphics[width=6cm]{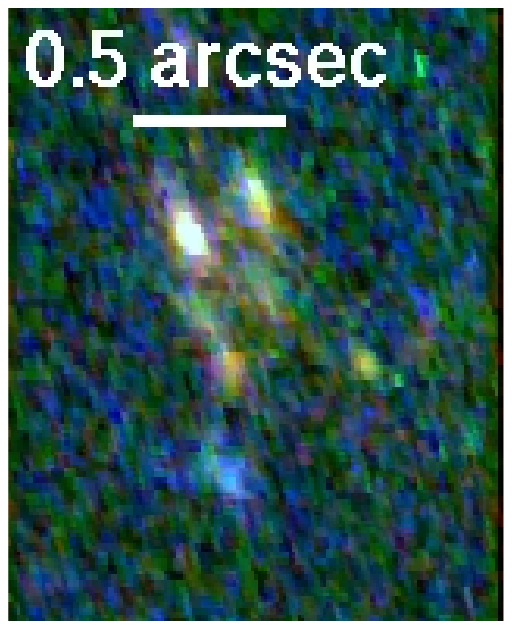}
\caption[]{Reconstruction of the source plane image of J163556 shown using
WCS images in the filters 625W, 775W, and 850LP. 
\label{fig:lens_reconstr} }
\end{center}
\end{figure}

Adopting a redshift of 4, we estimate the gravitational lensing magnification
using the cluster potential model from \citep{eliasdottir08} as input to
LENSTOOL \citep{jullo07}.  We find a magnification factor of  $5.5\pm1.9$ 
\footnote{We note that this value, although different, is consistent with the
magnification factor that we used in \citet{knudsen09}. The difference is a
consequence of a recent revision of the model in \citep{eliasdottir08}, which
affects this region of the cluster, and the fact that in
\citet{knudsen09} was determined for a single position, while here an
extensive analysis to determine the uncertainty was carried out. Differences
of 12\%-20\% are generally to be expected. }  
and note that this does not change much for redshift $z>2$.  
The spatial magnification is a factor 3.2 in the direction tangential to the
direction of the cluster center and 1.7 in the direction perpendicular. 
In Fig.~\ref{fig:lens_reconstr} we show the reconstruction of the source
plane image of J163556 using the WCS images.  The morphology of the object is
only slightly distorted, albeit all the scales are reduced by a factor of
$\sim 2$.  The different colors of the clumps are likely due to different
reddening. 

The CO(4-3) line was not detected for J163556 in IRAM Plateau de Bure
observations in the redshift range $z=4.035-4.082$ \citep{knudsen09}.
Assuming a line width of 400 km/s \citet{knudsen09} placed an upper limit
on the CO line luminosity of $0.3\times10^{10}$\,K\,km/s\,pc$^2$ (using a lensing
magnification of 5.5).  This upper limit is not inconsistent with the
$L'_{\rm CO} - L_{\rm FIR}$ relation for starburst galaxies\footnote{We note
that the far-infrared luminosity given in \citet{knudsen09} unfortunately
missed a factor $(1+z)^{-1}$.}, hence the non-detection does not contradict the
determined redshift $z=4.044$.  
It is important to keep in mind that we do not know if the CO(4-3) transition
is thermalized.  While several bright SMGs have been found to have
thermalized CO at least up to the (3-2) transition \citep[e.g.][]{tacconi08}, 
cases of Milky Way type molecular gas excitation conditions have been found
for a $z\sim1.5$ BzK galaxy \citep{daddi08,dannerbauer09} as well as
subthermal excitation of the CO(2-1) and CO(5-4) in the submm bright
extremely red object HR10 \citep{greve03,papadopoulos02}.  
The CO(4-3) non-detection could also be explained by
a velocity shift between the Lyman-$\alpha$ and the CO larger than
$-525$\,km/s, which is the difference between $z=4.035$ and $4.044$ for
CO(4-3).  Shifts of $z_{\rm Ly\alpha} - z_{\rm CO} = -500$\,km/s have
been seen for other SMGs \citep{greve05}.

\subsection{Spectral energy distribution}

\begin{figure}
\begin{center}
\includegraphics[width=8.5cm]{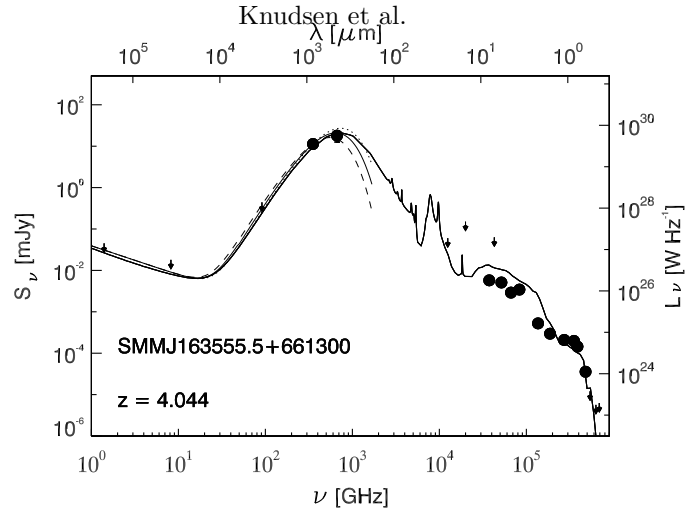}
\caption[]{The SED of J163556, overlayed with a modified blackbody with $T_d
= 30,35,40$\,K (dashed, solid, dotted, respectively) and $\beta=1.5$, as well
as the SED model for Arp220 from \citet{silva98} scaled to the submm points
of J163556.  
\label{fig:sed} }
\end{center}
\end{figure}

Based on the 850$\mu$m and the 450$\mu$m data points, the far-infrared SED is
well-described by a modified black-body spectrum, with a dust temperature
$\sim 35$\,K assuming $\beta = 1.5$, where $\beta$ is $f_{\nu} \propto
\nu^{\beta}$.  
The dust temperature is comparable to what is seen in other SMGs at lower 
redshift \citep[e.g.][]{kovacs06}. 
If the single optical emission line would not be Lyman-$\alpha$, but [OIII]
or [OII] implying a low redshift, the implied dust temperature would be
$9-11$\,K. 
In Fig.~\ref{fig:sed} we show the multiwavelength SED overlayed with a
modified blackbody spectrum and the Arp220 SED from \cite{silva98}. 
The far-infrared luminosity derived for this redshift 4.044 is
$1.3^{+0.7}_{-0.3}\times10^{12}$\,L$_\odot$ after correcting for
gravitational lensing and where the two uncertainties represent the upper and
lower limit from the magnification correction.  
J163556 is intrinsically a fainter object than the other $z>4$ SMGs, 
which have estimated FIR luminosities
$\sim10^{13}$\,L$_\odot$ \citep{capak08,daddi09,daddi09b,coppin09}. 
Assuming a Salpeter initial mass function, solar abundance, a continuous
burst for 10-100 Myr, and that the dust reradiates all the bolometric
luminosity, the far-infrared luminosity implies a star formation rate of
$\sim 230$\,M$_\odot$yr$^{-1}$ 
\citep{kennicutt98}. 

Assuming that the submm and radio emission from J163556 are correlated, we
use the radio-submm flux ratio as function of redshift as discussed by
\citet{blain99} to estimate the radio flux.  We estimate $S_{1.4} \sim
40$\,$\mu$Jy.  Assuming that the radio SED is described by a power law,
$S_{\nu}\propto \nu^{\alpha}$ with $\alpha = -0.7$, then $S_{8.2} \sim
13$\,$\mu$Jy.  Both estimated fluxes are below the sensitivity levels of the
WSRT and VLA maps.    This implies that no radio-loud AGN is present and thus
the submm and radio emission is most likely powered by star formation
rather than AGN activity.  The absence of a dominating AGN or quasar is also
supported by the lack of an X-ray detection in the {\em Chandra} observations
\citep{machacek02,govoni04}; in the $0.8-4$\,keV band an upper limit of
$1.24\times10^{-14}$\,erg/s/cm$^2$ is estimated for an area of $3\times3$
pixels.    
We note that this is the first $z>4$ SMG identification without radio
counterpart, although we also point out that the radio observations are not
as deep as those used for the identification of other $z>4$ SMGs. 

To determine the properties of J163556 we use {\it hyperz} and find that the
optical and near-infrared photometry is best fit by a 
star formation history of an elliptical galaxy template with an age of
$\sim 1.4$\,Gyr and a reddening of $A_V = 0.8$\,mag.  The rest-frame $K$-band
absolute magnitude is -24.40.  For this template the $K$-band luminosity
would correspond to a stellar mass of $8.7\times10^{10}$\,M$_\odot$ without
correcting for the gravitational lensing. Correcting for lensing
magnification this corresponds to $\sim1.6\times10^{10}$\,M$_\odot$. 
The results of the fit, however, yield an age close to the age of the
Universe at that redshift, $\sim1.5$\,Gyr.  Additionally, we fix the redshift
to the spectroscopic redshift and use different star formation histories from
a single burst to a continuous star formation.  A single burst gives the best
fit with a reduced $\chi^2$ of 1.39, a reddening of $A_V = 1.8$\,mag and very
young age of 10-20 Myrs.  Other star formation histories with an exponential
decay or with continuous star formation yield a $\chi^2$ of 1.76, reddening
of $A_V= 2.0$\,mag, and a best fit age of 40 Myrs.  

We note that the {\em Spitzer} IRAC observations probe the light at
rest-frame wavelengths of $0.71-1.6\mu$m, which for an elliptical galaxy is 
dominated by the evolved stellar populations.  However, for a starburst
galaxy this could be dominated by young supergiants, as is seen in local
ULIRGs.  It is known that there is a degeneracy between a mature stellar
population with small or modest amount of extinction and a younger population
with a much stronger extinction.  A younger population would imply a lower
stellar mass estimate and thus the best fit estimate can be considered an
upper limit. 

Based on the F775W filter, we can compute the rest-frame UV continuum
luminosity $L(1500{\mathrm \AA}) = 2.0\times10^{22}$\,W/Hz and derive an
unlensed star formation rate (SFR) of $\sim28$~M$_\odot$\,yr$^{-1}$ using the
\citet{kennicutt98} empirical calibration (although certainly underestimated due
to dust extinction).  This result compared to the infrared luminosity
suggests that the vast majority of the young stars in this galaxy are
obscured by dust and are undetectable in the rest-frame far-UV.

\subsection{Size and morphology}

The HST images give the best insight to the morphology of the galaxy.  The
resolution is the highest in the ACS F850LP ($z'$-band) image, which probes
redshifted UV light (see Fig.\ \ref{fig:phot}).  The galaxy is seen as a very
irregular system made of four knots.  The total extent of the galaxy in the
$z'$-band image is $1.8''\times 1.1''$. Correcting for the lensing
magnification, the source size is then $0.9''\times 0.4''$.  For $z=4.044$
this corresponds to a physical scale of $\sim6$\,kpc $\times$ 3\,kpc.  The
largest star forming knot is unresolved in the ACS images and and we place an
upper limit on the extent of about 0.07'' or 500\,pc in the source plane.
In the $K$-band (which corresponds to rest-frame $B$-band) image the light
originates primarily from the central part of the galaxy.  We suggest that
there already are old stars in place at the center of this galaxy. 

J163556 is detected as a point source in the SMA data.  Correcting the beam
shape for a gravitational lensing amplification in the of $3.2\times 1.7$ in
the tangential and the radial direction, respectively, of the cluster
potential well we find the beam to be $0.47''\times1.2''$ in the source plane.
This is equivalent to the resolution that can be obtained with the SMA in
extended configuration without gravitational lensing.  
Through a 2D-Gaussian fit to the SMA data we are not able to resolve the
source concluding that it is point-like within the SMA beam.  We place a
strict upper limit on the size of the starburst region using the lensing
corrected beam. 
In the $z=4.044$ source plane this
corresponds to $8{\rm kpc}\times3{\rm kpc}$. 
For a star formation rate of $\sim 230 $\,M$_\odot$\,yr$^{-1}$, this yields
a lower limit for the star formation rate density of
$12$\,M$_\odot$\,yr$^{-1}$\,kpc$^{-2}$.

\subsection{Extending the redshift distribution of faint SMGs}

The previously identified five $z>4$ bright SMGs have far-infrared
luminosities $5-10$ times larger than J163556.  For the six other reliably
identified faint SMGs with $S_{\rm 850\mu m} < 2$\,mJy \citep[][; Knudsen et
al., in prep]{frayer03,kneib04,borys04,knudsen09}, the redshifts range from
1.03 to 2.9 with majority of the sources found around $z\sim 2.5$.  Due to
the modest number of known faint SMGs, their redshift distribution is poorly
constrained and it has been proposed by several groups that the faint SMGs
or low luminosity SMGs peak at lower redshifts than the bright SMGs
\citep[e.g.][]{wall08}.  Our result for J163556 thus extends the redshift
distribution of faint SMGs to very high redshifts.  

\begin{figure}
\begin{center}
\includegraphics[width=8.5cm]{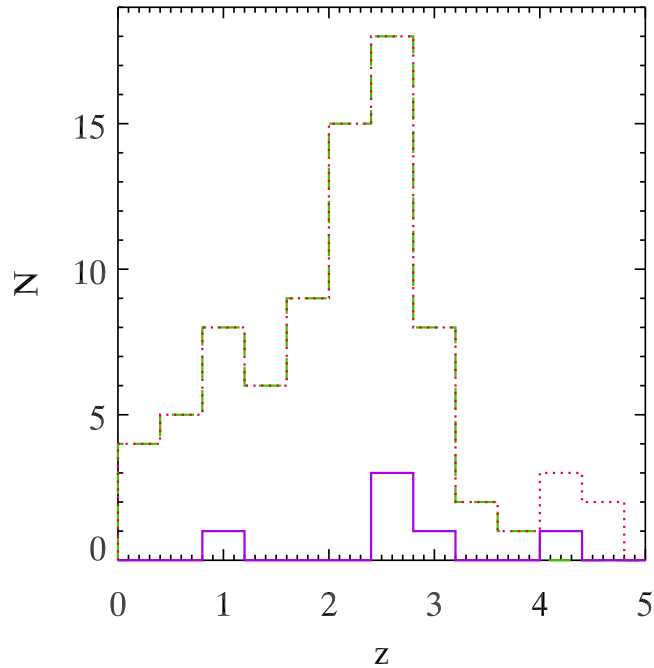}
\caption[]{The redshift distribution for faint and bright SMGs.  The solid
(purple) histogram shows the redshift distribution of the reliably identified
lensed SMGs with $S_{\rm 850\mu m} < 2$\,mJy.  The long dashed (green)
histogram shows the redshift distribution from \citet{chapman05}, and the
dotted (red) histogram shows the same including the five redshifts from
\citet{capak08,coppin09,daddi09,daddi09b}.  
\label{fig:z_dist} }
\end{center}
\end{figure}

In Fig.~\ref{fig:z_dist} we plot the redshift distribution of $S_{\rm 850\mu
m} < 2$\,mJy SMGs with spectroscopic redshifts in comparison to the redshift
distribution for radio identified, blank field SMGs \citep{chapman05}.
Between redshift 1 and 8, due to the negative $k$-correction in the submm,
the observed flux density stays essentially constant for a given far-infrared
luminosity. As several SMGs have their far-infrared luminosity determined
from single submm fluxes, we use the flux density rather than the luminosity
for our simple sample comparison. 
As the radio identified SMGs are expected to represent the general SMG
population, but with a possible bias against redshifts $z>3.5$, we also
include the five $z>4$ bright SMGs \citep{capak08,coppin09,daddi09,daddi09b}.
Using the non-parametric  rank sum test known as the Mann-Whitney U test
\citep[e.g.][]{babu96}, we
find that there is only a $1.3\sigma$ ($1.6\sigma$) level of significance
that the faint SMGs have a different redshift distribution than the Chapman
et al.\ sample + the $z>4$ SMGs (the Chapman et al.\ sample alone). 
Removing the $z<1$ tail of the Chapman et al.\ sample, as the negative
$k$-correction does not apply in that redshift range, we find a consistent
result. 
In other words, based on this small number of faint SMGs we find that it is
possible that the less luminous SMGs have a similar redshift distribution as
the more luminous SMGs.

With the model predictions for the redshift distribution of SMGs 
\citep[e.g.][]{chapman05,wall08} in most submm surveys (e.g.,
\citealt{smail02,webb03,coppin06,knudsen08}) at least a few sources are
expected to have a $z>4$.  As $z\sim 2$ massive, old evolved galaxies
($>10^{11}$\,M$_\odot$, $>2$\,Gyr) are observed with a space density of $\sim
10^{-4}$\,Mpc$^{-3}$ \citep[e.g.][]{daddi05,kong06}. These galaxies probably
had their dominant stellar population formed in a single starburst at
redshift $z>4$ \citep[e.g.][]{daddi05,stockton08}.  
The most likely progenitor of these are
the massive starbursts that we observe in SMGs.  If the large SFRs of $\sim
1000$\,M$_\odot$\,yr$^{-1}$ can be sustained over a period of $\sim100$\,Myr
it is possible to build such large stellar masses \citep[e.g.][]{coppin09}. 
Thus identification of $\geq 4$ SMGs is vital for our understanding of
$z\sim2$ massive, old galaxies. 

To reliably identify $z>4$ SMGs, deep data at
all wavelengths   is necessary.  Though the strongest, initial probe comes
from the ratio between 450\,$\mu$m and 850\,$\mu$m fluxes, or possibly
350\,$\mu$m and 850\,$\mu$m, where this ratio is expected to be $<2$.  In
most cases this requires very deep 450\,$\mu$m or 350\,$\mu$m observations
with a 1$\sigma$ rms $<5$\,mJy/beam.  Thus the possibilities for systematic
searches for very high redshift SMGs rely on the success of the future
high-sensitive submillimeter cameras such as SCUBA-2 
and space telescopes such as Herschel to provide such observations.

\section{Conclusions} 

We report the identification of the gravitationally lensed SMG
SMMJ163555.5+661300 (J163556), which has a lensing corrected far-infrared
luminosity of $1.3\times10^{12}$\,L$_\odot$.  Using SMA interferometric
observations, we have identified the underlying optical counterpart.  The SMA
flux agrees well with the SCUBA flux of 11.3 mJy.  
\begin{itemize}
\item From optical spectroscopy from the Keck Telescope, we detect an asymmetric
emission line at $6135$\,\AA\ and no emission on the blue side.  We identify
this emission line as the Lyman$\alpha$ line, which implies a redshift of
$4.0441\pm0.001$.  Using photometric redshift calculations we rule out that
the single emission line could be [OII] or [OIII], which would otherwise have
implied a redshift of 0.25 and 0.61, respectively.  This is also strongly
supported by the flux ratios at submm and radio wavelengths.  
\item J163556 is not resolved in the SMA data, and we place a strict upper
limit on the starburst size of $8{\rm kpc}\times 3{\rm kpc}$.  From the
derived SFR of 230\,M$_\odot$\,yr$^{-1}$ we estimate a lower limit on the SFR
surface density of $12$\,M$_\odot$\,yr$^{-1}$\,kpc$^{-2}$.  
\item
With an unlensed flux of $\sim2$\,mJy, J163556 is fainter than the classical
(bright) SMGs that are often classified at $S_{\rm 850\mu m} > 5$\,mJy.  It
is thus so far the most distant known faint SMG.  This result extents the
redshift distribution of the faint SMGs that are typically only found in
gravitational lensing surveys to look-back times equivalent to the bright
SMGs.  
\item  The derived upper limit for the CO(4-3) line luminosity of
$0.3\times10^{10}$\,K km/s pc$^{-2}$ is not inconsistent with the $L^{'}_{\rm
CO} - L_{\rm FIR}$ relation for starburst galaxies.  The CO(4-3)
non-detection is possibly caused by subthermal excitation or a large velocity
shift.  
\item  Based on the small number of reliably identified faint, lensed SMGs
with spectroscopic redshift, we do not find evidence that they have a
different redshift distribution than that of bright SMGs as traced by the
radio identified SMGs from \citet{chapman05}.  

\end{itemize}
\


%% Included in this acknowledgments section are examples of the
%% AASTeX hypertext markup commands. Use \url without the optional [HREF]
%% argument when you want to print the url directly in the text. Otherwise,
%% use either \url or \anchor, with the HREF as the first argument and the
%% text to be printed in the second.

\acknowledgments

We thank an anonymous referee for insightful comments, which help improve the
manuscript.  
The authors recognize and acknowledge the very significant cultural role and
reverence that the summit of Mauna Kea has always had within the indigenous
Hawaiian community. We are most fortunate to have the opportunity to conduct
observations from this mountain.  This program is based on observations made
with the NASA/ESA Hubble Space Telescope, which is operated by the
Association of Universities for Research in Astronomy, Inc., under NASA
contract NAS 5-26555, the Subaru Telescope, which is operated by the National
Astronomical Observatory of Japan, the Spitzer Space Telescope, which is
operated by the Jet Propulsion Laboratory, California Institute of Technology
under a contract with NASA.
The Submillimeter Array is a joint project between the Smithsonian
Astrophysical Observatory and the Academia Sinica Institute of Astronomy and
Astrophysics and is funded by the Smithsonian Institution and the Academia
Sinica.
We thank Federica Govoni and Marie Machacek for providing the Chandra X-ray
upper limit.  
KK acknowledges support from the Deutsche Forschungsgemeinschaft (DFG)
Priority Programme 1177.  JPK thanks CNRS for support.
JR acknowledges support from a EU Marie-Curie fellowship.


%% copy editing.  Individual instruments can be provided in parentheses,
%% after the keyword, but they will not be verified.

{\it Facilities:} \facility{SMA}, \facility{Keck (LRIS)}, \facility{JCMT
(SCUBA)}, \facility{HST (ACS, NICMOS)}, \facility{Spitzer (IRAC, MIPS)},
\facility{WSRT}, \facility{VLA}, \facility{Subaru (MOIRCS)}




%\bibliography{apj-jour,z4.bib}

\clearpage

\begin{deluxetable}{llccc}
\tablecolumns{5}
\tabletypesize{\footnotesize}
\tablewidth{0pc}
\tablecaption{Photometry for J163556. \label{tab:phot} }
\tablehead{
\colhead{Telescope} & \colhead{$\lambda$} & \colhead{flux} & \colhead{R.A.(J2000)} & \colhead{Dec.(J2000)} \\
}
\startdata
WSRT& 1.4\,GHz &  $<45$\,$\mu$Jy && \\
VLA & 8.2\,GHz &  $<18$\,$\mu$Jy && \\
IRAM & 91\,GHz &  $<0.44$\,mJy && \\
SMA & 870\,$\mu$m & $10.7\pm2.0$\,mJy & 16:35:55.67 & +66:12:59.51 \\  %from Miriad uvfit
JCMT & 850\,$\mu$m & $11.3\pm1.3$\,mJy & 16:35:55.5 & +66:13:00 \\
JCMT & 450\,$\mu$m & $18.0\pm5.4$\,mJy &&\\
Spitzer & 24.0\,$\mu$m & $<60$\,$\mu$Jy && \\
ISO & 15\,$\mu$m & $<150$\,$\mu$Jy && \\
Spitzer & 8.0\,$\mu$m & $22.03\pm0.10$\,mag && \\
ISO & 6.7\,$\mu$m & $<63$\,$\mu$Jy && \\
Spitzer & 5.8\,$\mu$m & $22.16\pm0.10$\,mag  &&\\
Spitzer & 4.5\,$\mu$m & $22.77\pm0.05$\,mag  && \\
Spitzer & 3.6\,$\mu$m & $22.59\pm0.05$\,mag  & &  \\ 
Subaru & $K$ & $24.63\pm0.06$\,mag & 16:35:55.64 & +66:12:59.32 \\
HST & F160W & $25.25\pm0.05$\,mag & 16:35:55.66 & +66:12:59.45 \\
HST & F110W & $25.63\pm0.06$\,mag & 16:35:55.64 & +66:12:59.37 \\
HST & F850LP & $25.69\pm0.05$\,mag & 16:35:55.69 & +66:12:59.49 \\
HST & F775W & $26.03\pm0.05$\,mag & 16:35:55.69 & +66:12:59.52 \\
HST & F625W & $27.55\pm0.13$\,mag & 16:35:55.71 & +66:12:59.34 \\
%  mean position for the optical + nir: 16:35:55.67±0.03, +66:12:59.42±0.08
HST & F555W & $>28.73$\,mag && \\
HST & F475W & $>29.55$\,mag & & \\
HST & F435W & $>29.44$\,mag & &  \\
Chandra & $0.8-4$\,keV & $<1.24\times10^{-14}$ erg/s/cm$^2$ & & 
\enddata
\end{deluxetable}

\end{document}